\documentclass[structabstract]{aa}
\usepackage{graphicx}
\usepackage{txfonts}
\usepackage{natbib}

\begin{document}

\title{Spot activity of \object{LQ~Hya} from photometry between 1988 and 2011
\thanks{The analysed photometry and numerical results of the analysis are both published electronically at the CDS via anonymous ftp to cdsarc.u-strasbg.fr (130.79.128.5) or via http://cdsarc.u-strasbg.fr/viz-bin/qcat?J/A+A/yyy/Axxx}}
\author{J. Lehtinen \inst{1,2}
\and L. Jetsu \inst{1}
\and T. Hackman \inst{1,3}
\and P. Kajatkari \inst{1}
\and G.W. Henry \inst{4}}
\institute{Department of Physics, Gustaf H\"{a}llstr\"{o}min katu 2a (P.O. Box 64), FI-00014 University of Helsinki, Finland
\and Nordic Optical Telescope, 38700 Santa Cruz de La Palma, Spain
\and Finnish Centre for Astronomy with ESO, University of Turku, 
V\"{a}is\"{a}l\"{a}ntie 20, FI-21500 Piikki\"{o}, Finland
\and Center of Excellence in Information Systems, 
Tennessee State University, 3500 John A. Merritt Blvd., Box 9501, Nashville, TN 37209, USA}
\date{Received / Accepted}

\abstract{}
{We investigate the spot activity of the young magnetically active main sequence star \object{LQ~Hya}. Our aims are to identify possible active longitudes, estimate the differential rotation and study long and short term changes in the activity.}
{Our analysis is based on 24 years of Johnson V-band photometry of \object{LQ~Hya} obtained with the T3 0.4m Automated Telescope at the Fairborn Observatory. We use the previously published Continuous Period Search (CPS) method to model the evolution of the light curve of \object{LQ~Hya}. The CPS fits a Fourier series model to short overlapping subsets of data. This enables us to monitor the evolution of the light curve and thus the spot configuration of the star with a higher time resolution.}
{We find seasonal variability in the mean level and amplitude of the light curve of \object{LQ~Hya}. The variability of the light curve amplitude seems not to be cyclic, but the long-term variations in the mean magnitude may be indicative of an approximately 13 year cycle. However, because of the limited length of the observed time series, it is not yet possible to determine whether this structure really represents an activity cycle. Based on fluctuations of the light curve period, we estimate the differential rotation of the star to be small, and the star is potentially very close to a rigid rotator. We search for active longitudes from the inferred epochs of the light curve minima. We find that on time scales up to six months there are typically one or two relatively stable active areas on the star with limited phase migration. On the other hand, on time scales longer than one year, no stable active longitudes have been present except for the period between 2003 and 2009 and possibly also some time before 1995. Neither do we find any signs of flip-flops with a regular period. The mean time scale of change of the light curve during the observation period is determined to be of the same order of magnitude as the estimated convective turnover time for the star.}
{}

\maketitle

\section{Introduction}

\object{LQ~Hya} (\object{HD~82558}) is a young single magnetically active star ($V=7.8$, $B-V=0.9$, K2V) classified as a BY Dra star by \citet{fekel1986survey}. Its strong activity is clearly evident from the substantial Ca HK emission, $\log{R'_{\rm HK}}=-4.06$ \citep{white2007high}, placing it confidently within the ``very active'' regime defined by \citet{henry1996survey}.

It was suggested by \citet{fekel1986chromospherically} that the star is a very young object which has recently arrived at the zero-age main-sequence. \citet{montes2001late} classified it as a member of the young disc population following the definition of \citet{eggen1984a0,eggen1989large}. Recently \citet{nakajima2012potential} identified the star as a possible member of the \object{IC~2391} supercluster, thus estimating its age to be 35--55 Myrs \citep{montes2001late}.

The magnetic activity of \object{LQ~Hya} is strongly manifested as starspots causing rotational modulation of brightness \citep{eggen1984systematic}. The rotation period of the star can be measured from this modulation. Previously the rotation period of \object{LQ~Hya} has been estimated as $P = 1.601136 \pm 0.000013$ d by \citet{jetsu1993decade}, $P = 1.601052 \pm 0.000014$ d by \citet{berdyugina2002magnetic} and $P = 1.60066 \pm 0.00013$ d by \citet{kovari2004doppler}. Some variability of the observable photometric rotation period is expected, however. This can be caused by the spots moving with different angular velocities governed by the underlying surface differential rotation or large scale magnetic field. Also changes in the light curve shape governed by active regions growing and decaying at different locations on the stellar surface may cause photometric variations unrelated to spot rotation.

The differential rotation of \object{LQ~Hya} is found by many authors to be small. \citet{jetsu1993decade} analysed the fluctuations of the photometric period of \object{LQ~Hya} within $3\sigma$ limits and reported the relative scale of them to be $Z \approx 0.015$. This value can be interpreted as the relative scale of rotation periods of the observed starspots. Provided that the spots trace the surface rotation of the star and that there have been spots on all stellar latitudes from the equator to the poles, we may estimate the differential rotation coefficient to be $k \approx 0.015$. \citet{you2007photometric} used a similar approach to estimate the differential rotation of the star from the light curve period fluctuations and derived the value $k \approx 0.025$. Both of these values are similar to the theoretical estimate $k=0.0128$ obtained for \object{LQ~Hya} by \citet{kitchatinov2011differential} using a mean field hydrodynamical model.

Even smaller differential rotation values were reported by \citet{berdyugina2002magnetic} and \citet{kovari2004doppler} who both used Doppler images of \object{LQ~Hya} in conjunction with photometry in their analyses. \citet{berdyugina2002magnetic} compared photometric periods of \object{LQ~Hya} to Doppler images corresponding to the same epoch of time to infer the latitude of the main spot. Using this approach they retrieved a differential rotation coefficient of $k \approx 0.002$. \citet{kovari2004doppler} used cross correlation between adjacent Doppler images to estimate the differential rotation and reported $k=0.0057$.

Many authors have reported signs of cyclic behaviour in the spot activity of \object{LQ~Hya}. \citet{jetsu1993decade} found a 6.24 yr cycle in the mean brightness of the star using light curve fits for 9 years of photometric observations. \citet{olah2000multiperiodic} applied a Fourier analysis to 16 years of photometry and found cycle periods of both 6.8 yr and 11.4 yr in the mean brightness. Based on light curve inversions from 20 years of photometry, \citet{berdyugina2002magnetic} reported a 15 yr cycle for the mean brightness, a 7.7 yr cycle for the light curve amplitude modulation and a 5.2 yr flip-flop cycle. They also identified a 1:2:3 resonance between the cycle frequencies. \citet{kovari2004doppler} used Fourier analysis to search for cycles in the light curve mean brightness from 8 years of photometry and found a possible cycle of 13.8 yr along with its first harmonic 6.9 yr and also weak signs for a 3.7 yr cycle. Finally \citet{olah2009multiple} applied time-frequency analysis for 25 years of photometry of \object {LQ~Hya} and foud two short cycles of 2.5 yr and 3.6 yr as well as a longer cycle with its period increasing from 7 yr to 12.4 yr within the duration of the dataset.

In addition to \citet{berdyugina2002magnetic}, also \citet{jetsu1993decade} reported active longitudes for \object{LQ~Hya}. The nature of the active longitudes was, however, quite different. Where \citet{berdyugina2002magnetic} reported two active longitudes with $\Delta\phi=0.5$ phase separation, \citet{jetsu1993decade} claimed the phase separation between the active longitudes to be only $\Delta\phi=0.25$.

Doppler images of \object{LQ~Hya} have been reconstructed by \citet{strassmeier1993surface}, \citet{rice1998doppler}, \citet{donati1999magnetic}, \citet{donati2003dynamo} and \citet{kovari2004doppler}. They typically show spotted areas at mid latitudes relatively far from the visible pole. From time to time, there have been longitudinal concentrations of spots in these maps, but no clear pattern of stable active longitudes. Occasionally the surface reconstructions have shown a complete latitudinal band of spots encircling the star. On the other hand, reconstructions of the surface magnetic field using Zeeman Doppler imaging \citep{donati1999magnetic,donati2003dynamo} have sometimes shown opposite magnetic polarities on different sides of the visible pole. This may indicate at least an occasional presence of active longitude like features on the star.

\section{Analysis of the data}

\begin{figure}
\resizebox{\hsize}{!}{\includegraphics{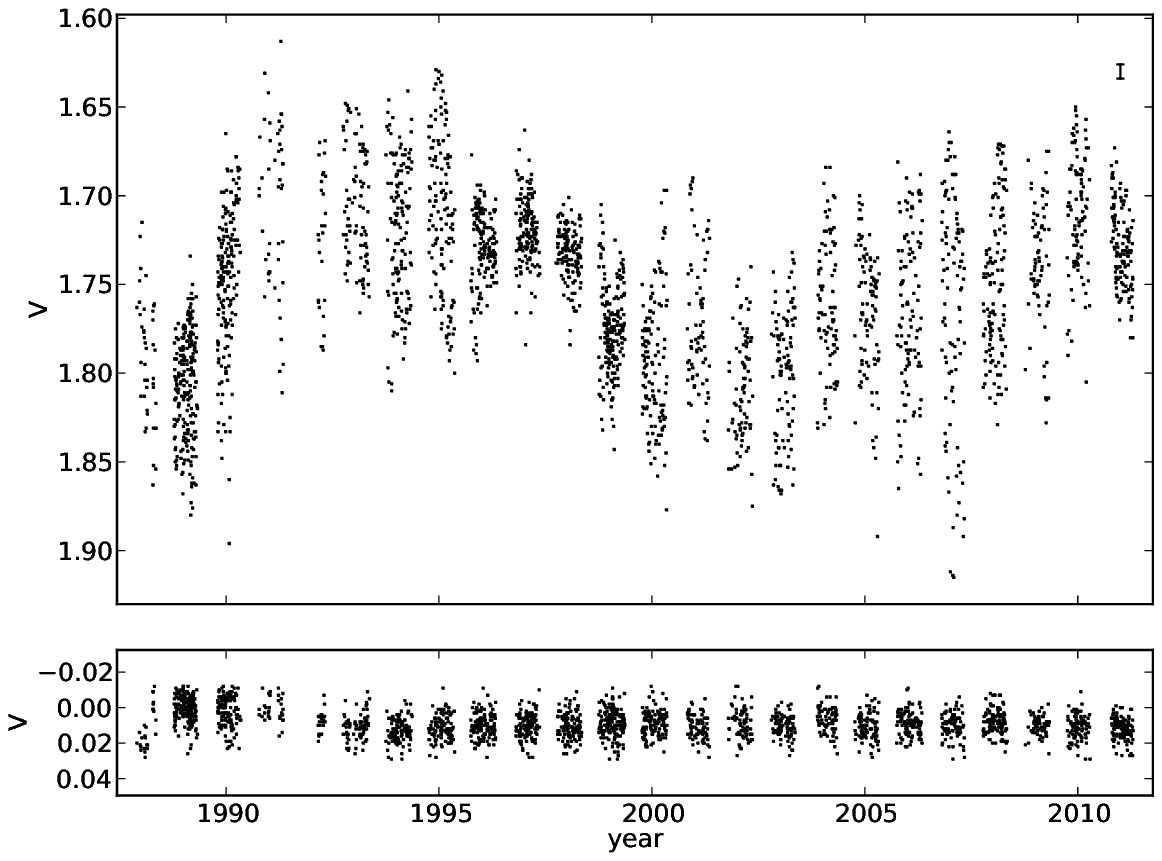}}
\caption{\textit{Top:} V band differential photometry of \object{LQ~Hya} minus comparison star. The small error bar in the upper right corner denotes the scale of the photometric uncertainty of $\pm 0.004$ mag. \textit{Bottom:} V band differential photometry of the check star minus comparison star. The scale is the same in both of the panels.}
\label{vdata}
\end{figure}

\begin{figure*}
\centering
\includegraphics[angle=90,width=17cm]{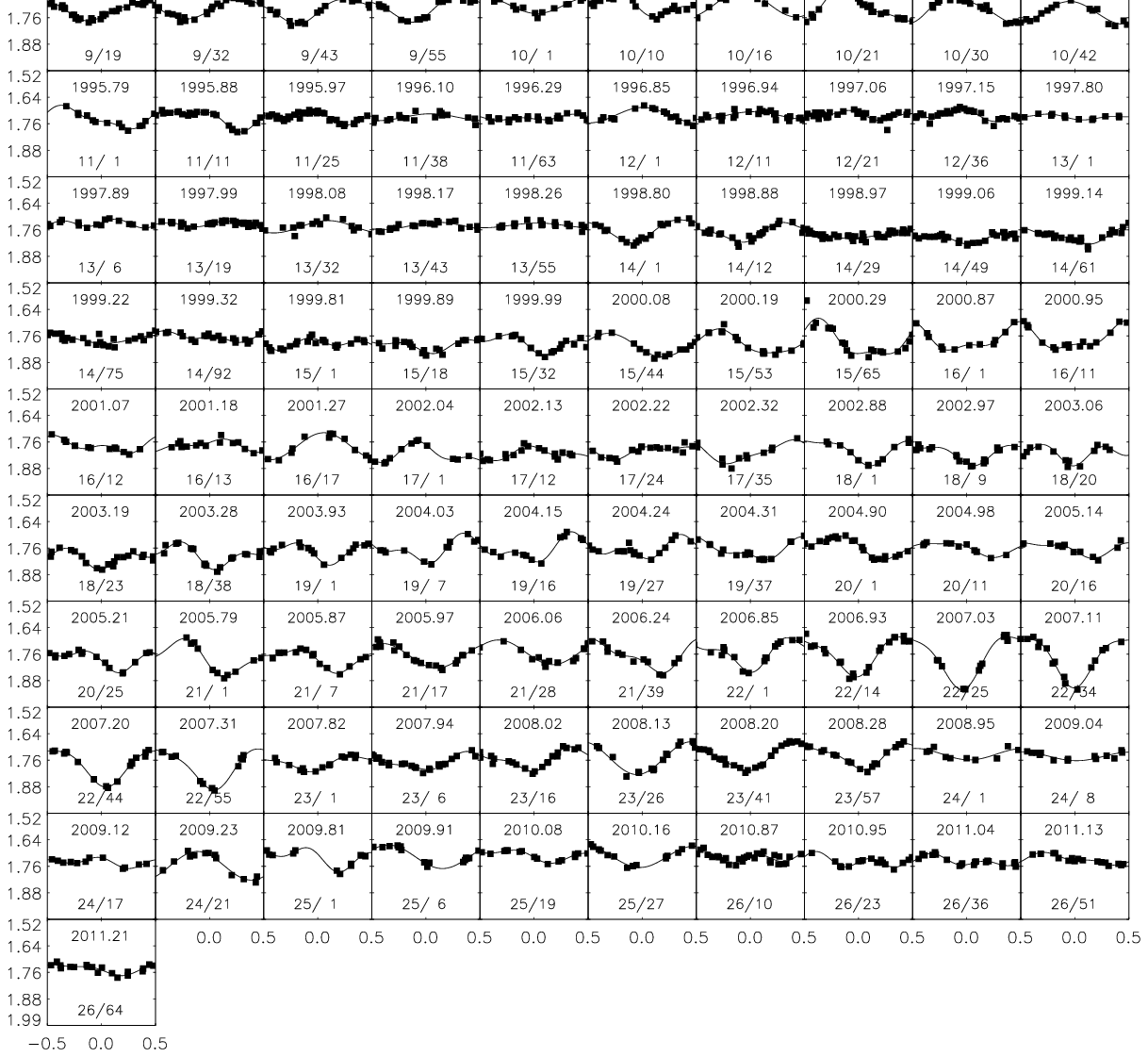}
\caption{Light curves and light curve fits of the 111 independent datasets. The ephemerides for the light curves are explained in Sect. \ref{ressect}. Each dataset is labelled with its mean epoch in year and the $\rm SEG/SET$ number.}
\label{allcurves}
\end{figure*}

The photometry of \object{LQ~Hya} was obtained over a 24 year time span between $\rm HJD=2447141$ (11 December 1987) and $\rm HJD=2455684$ (3 May 2011) with the T3 0.4 m Automatic Photoelectric Telescope (APT) at the Fairborn Observatory in Arizona. The complete dataset analysed in this paper consists of 2671 Johnson V-band observations of \object{LQ~Hya} minus the comparison star \object{HD~82477} ($V=6.13$, $B-V=1.19$). To monitor the constancy of the comparison star, 2272 additional simultaneous V-band observations of the check star \object{HD~82428} ($V=6.14$, $B-V=0.24$) were obtained.

The typical error of the target star photometry is estimated to be between 0.003 and 0.004 magnitudes from observations of constant stars \citep{henry1995development}. The errors of the check star observations are expected to be somewhat larger as fewer individual integrations have been used to determine their values. For a brief description of the operation of the APT and reduction of the data, see \citet{fekel2005chromospherically} and references therein.

Our Johnson V band \object{LQ~Hya} minus comp star and check star minus comp star differential magnitudes are presented in the top and bottom panels of Fig. \ref{vdata}, respectively. \object{LQ~Hya} shows significant variability on night-to-night (rotational), year-to-year (spot lifetime), and decadal (magnetic cycle) time scales. The check star observations, plotted on the same scale as the \object{LQ~Hya} observations show no evidence for variability on any time scale in either the comp star or the check star. The somewhat larger scatter and slight offsets seen in the first few years of the check minus comp star observations are the result of instrumental upgrades that affect mainly the check minus comp differential magnitudes because of the large color difference ($\Delta(B-V)=-0.95$) compared to LQ Hya minus comp ($\Delta(B-V)=-0.29$).

The photometry of \object{LQ~Hya} was analysed using the Continuous Period Search (hereafter CPS) method \citep{lehtinen2011continuous}. The method models the photometry in short subsets of the data by fitting a low $K$th order Fourier series,

\begin{equation}
\hat{y}(t_i) = M + \sum_{k=1}^K{\left[B_k\cos{(k2\pi ft_i)} + C_k\sin{(k2\pi ft_i)}\right]},
\label{model}
\end{equation}

\noindent to each dataset. This modelling provides estimates for the mean differential magnitude $M$, peak-to-peak amplitude $A$ and period $P=f^{-1}$ of the light curve, as well as epochs for the primary and a possible secondary minima $t_{\rm min,1}$ and $t_{\rm min,2}$. The parameters $M$ and $P$ are obtained directly from the parameters in Eq. \ref{model}, whereas $A$, $t_{\rm min,1}$ and $t_{\rm min,2}$ are determined numerically. The error and reliability estimates for the parameters are obtained from their bootstrap samples. To allow for variability in the model complexity, the CPS performs fits using models of orders $K=0\ldots2$. The model order $K=0$ corresponds to a simple constant brightness model and describes the absence of any intrinsic variability in the light curve. For each dataset, the Bayesian information criterion is applied to determine the best modelling order \citep[Eq. 6]{lehtinen2011continuous}.

The period search was done within a $\pm5\%$ interval around the a priori period estimate $P_0$. The search was limited to this interval because of the risk for interference with spurious periods. In line with the previous results from \object{LQ~Hya} we chose the value of $P_0=1.6$ d. The upper limit for length of the individual analysed datasets was set at $\Delta T_{\rm max}=30.4$ d, which is 19 times the length of $P_0$. This dataset length was chosen so that, even during times of sparse data sampling, most datasets would have enough data for modelling. On the other hand it is not too long to let the light curve shape change too much during the individual subsets. The choice of setting $\Delta T_{\rm max}$ as a integer multiple of $P_0$ was a precaution against uneven phase sampling of the light curve. To get good time resolution for the evolution of the light curve parameters $[M,A,P,t_{\rm min,1},t_{\rm min,2}]$, the datasets overlap with each other so that a new dataset was selected after 1 d from the start of the last one. Only datasets containing $n\geq12$ observations were included in the analysis. The mean of the residuals of all the model fits is $\overline{\epsilon}=0.009$.

\section{Results}
\label{ressect}

\begin{table}
\caption{Summary of the segments:
 Dates of the first and last analysed data point in HJD and calendar dates (yy/mm/dd), segment length $\Delta T_{\rm SEG}$ rounded into full days, the number of individual datasets $n_{\rm SET}$ and the number of mutually independent datasets $n_{\rm IND}$.}
\center
\begin{tabular}{llllll}
\hline
\hline
SEG \hspace{-3mm} & $\rm HJD-2400000$ \hspace{-1mm} & date & $\Delta T_{\rm SEG}$ \hspace{-2mm} & $n_{\rm SET}$ \hspace{-2mm} & $n_{\rm IND}$ \\
\hline
1 & 47199 -- 47230 & 88/02/08 -- 88/03/10 & 31 & 2 & 1 \\
2 & 47277 -- 47304 & 88/04/26 -- 88/05/23 & 27 & 1 & 0 \\
3 & 47460 -- 47660 & 88/10/25 -- 89/05/14 & 201 & 69 & 6 \\
4 & 47832 -- 48027 & 89/11/02 -- 90/05/16 & 195 & 47 & 5 \\
6 & 48348 -- 48394 & 91/04/02 -- 91/05/18 & 46 & 7 & 1 \\
7 & 48696 -- 48759 & 92/03/07 -- 92/05/17 & 63 & 6 & 1 \\
8 & 48911 -- 49132 & 92/10/15 -- 93/05/25 & 222 & 32 & 4 \\
9 & 49290 -- 49499 & 93/10/30 -- 94/05/27 & 209 & 60 & 6 \\
10 & 49645 -- 49866 & 94/10/19 -- 95/05/29 & 222 & 49 & 6 \\
11 & 50006 -- 50226 & 95/10/15 -- 96/05/23 & 221 & 67 & 5 \\
12 & 50391 -- 50582 & 96/11/04 -- 97/05/14 & 191 & 54 & 4 \\
13 & 50736 -- 50955 & 97/10/14 -- 98/05/22 & 220 & 61 & 6 \\
14 & 51103 -- 51325 & 98/10/16 -- 99/05/27 & 223 & 93 & 7 \\
15 & 51474 -- 51687 & 99/10/23 -- 00/05/23 & 213 & 69 & 6 \\
16 & 51861 -- 52052 & 00/11/13 -- 01/05/23 & 191 & 20 & 5 \\
17 & 52287 -- 52421 & 02/01/12 -- 02/05/27 & 135 & 36 & 4 \\
18 & 52594 -- 52785 & 02/11/15 -- 03/05/23 & 192 & 46 & 5 \\
19 & 52977 -- 53149 & 03/12/04 -- 04/05/24 & 172 & 37 & 5 \\
20 & 53329 -- 53506 & 04/11/20 -- 05/05/16 & 177 & 37 & 4 \\
21 & 53660 -- 53876 & 05/10/16 -- 06/05/21 & 217 & 43 & 5 \\
22 & 54044 -- 54238 & 06/11/05 -- 07/05/18 & 194 & 55 & 6 \\
23 & 54400 -- 54599 & 07/10/26 -- 08/05/13 & 200 & 57 & 6 \\
24 & 54810 -- 54966 & 08/12/10 -- 09/05/15 & 156 & 26 & 4 \\
25 & 55126 -- 55301 & 09/10/21 -- 10/04/15 & 176 & 31 & 4 \\
26 & 55499 -- 55684 & 10/10/29 -- 11/05/03 & 186 & 72 & 5 \\
\hline
\end{tabular}
\label{segtab}
\end{table}

The CPS automatically divides the data into segments consisting of mutually overlapping datasets. In addition to this, it cleans the data by removing outliers and temporally isolated data points \citep[see][]{lehtinen2011continuous}. The segment division for the data of \object{LQ~Hya} is summarised in Table \ref{segtab}. The table lists the dates of the first and last analysed data point in each segment, the lengths of the segments in days, as well as the numbers of all analysed datasets and independent datasets in the segments. The provisional segment SEG 5 is lacking from the listing because it did not contain enough data for modelling, i.e. it contained only some isolated data points. Overall, the data from the first few years is fragmentary, resulting in some shorter segments.

The mutually independent datasets were selected from all datasets so that they do not overlap with each other. Using the complete set of analysed datasets gives a detailed view into the time evolution of the light curve. On the other hand, using only the independent datasets removes any effects introduced by mutual correlations between the models of partially overlapping datasets.

The numerical results for all of the 1077 analysed datasets can be accessed electronically at the CDS. Light curve fits for the 111 independent datasets are presented in Fig. \ref{allcurves}. Observations in the datasets have been folded according to $\phi={\rm FRAC}[(t-t_{\rm min,1})/P]+\phi_{\rm min,1}$, where ${\rm FRAC}[x]$ removes the integer part of $x$ and $\phi_{\rm min,1}$ is the modelled light curve minimum phase computed using the ephemeris of Eq. \ref{ephem}. The use of two different periods here is necessary to both preserve the internal phase structure of the light curves and to visualise the long term phase evolution of the light curve minima. The ephemeris of Eq. \ref{ephem} is defined later in Sect. \ref{alsect} in the context of active longitudes.

\subsection{Long term variability of $M$, $A$ and $P$}

\begin{figure}
\resizebox{\hsize}{!}{\includegraphics{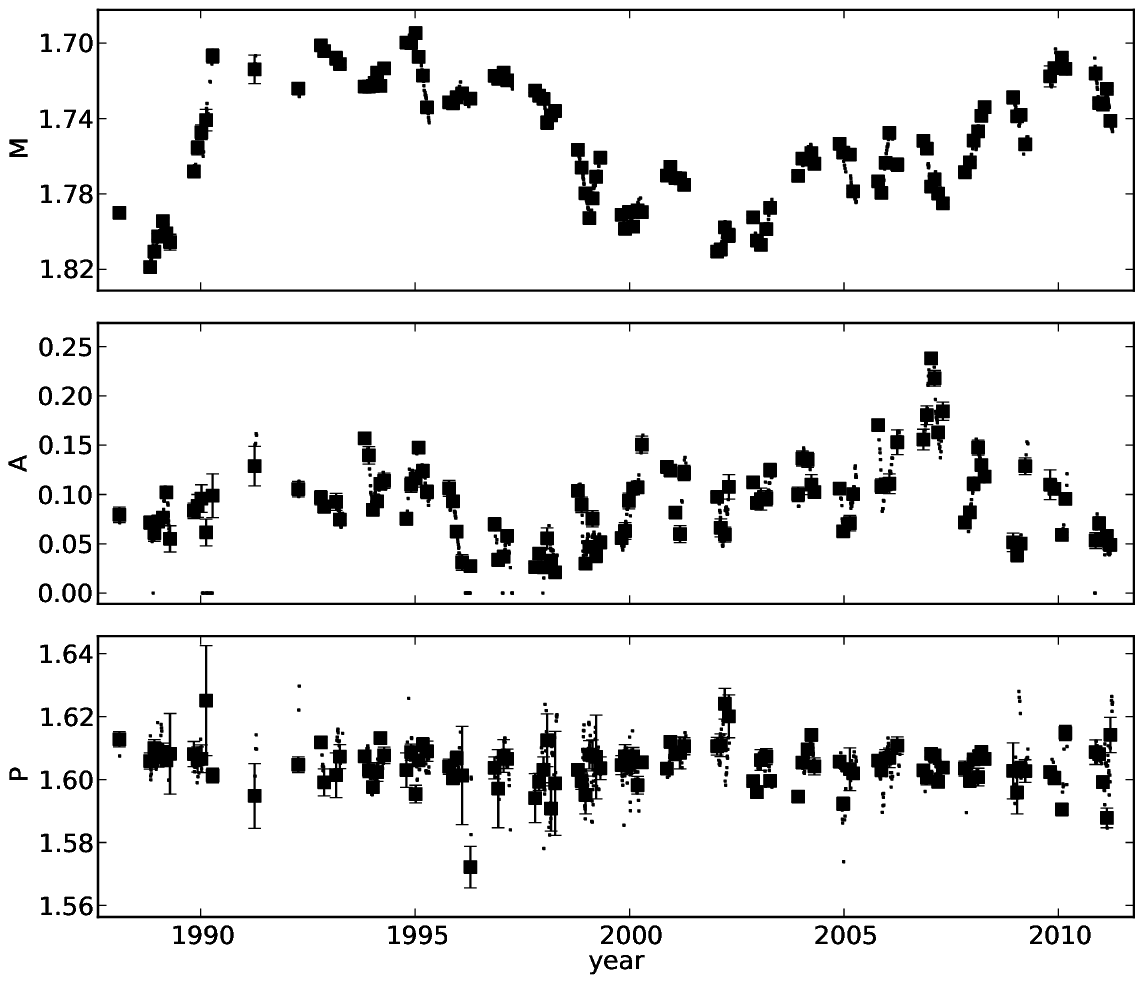}}
\caption{Long term variation of the mean differential magnitude $M$ (\textit{top}), total light curve amplitude $A$ (\textit{middle}) and photometric period $P$ (\textit{bottom}). $M$ and $A$ are given in magnitudes and $P$ in days. Squares with error bars indicate parameter estimates from the independent datasets and dots parameter estimates from all other reliable datasets.}
\label{cyc}
\end{figure}

The long term evolution of the light curve mean $M$, amplitude $A$ and period $P$ are presented in Fig. \ref{cyc}. We find both regular trends and random fluctuations from these light curve parameters during the last 24 years of observations.

The regular variations are most striking in the light curve mean $M$ which naturally follows the overall variation of the raw photometry (Fig. \ref{vdata}). The trend in the $M$ variations resembles what may turn out to be a regular activity cycle. There was a brightness minimum at 1989 and again around 2002. Between these two minima the mean brightness of the star increased with an amplitude of nearly 0.1 mag. After the 2002 minimum the mean brightness has climbed steadily towards a new maximum.

The variations in $M$ could be explained with an activity cycle of approximately 13 years. However, the total length of observations only includes what seems to be one and a half cycles, so conclusively proving the existence of such a cycle, let alone accurately determining its length, remains unfeasible. Such considerations are rendered even harder due to the fact that the profiles of these two apparently separate cycles in $M$ are quite different. What may appear as cyclic behaviour in the light of the available data may turn out to be random variability in the future and vice versa. It should be noted, however, that both \citet{berdyugina2002magnetic} and \citet{olah2009multiple} included photometry of \object{LQ~Hya} in their papers going back to November 1982. This additional data shows that the mean brightness of the star was decreasing during the years preceding the start of our photometry and thus fits qualitatively to the idea of a 13 yr cycle.

Also another possible short time scale cyclic pattern can be seen in the $M$ estimates, especially after 2005. These variations have had a period of around 2 years and amplitude around 0.02 mag. They may well have been responsible for the short 2.5 yr and 3.6 yr cycles reported by \citet{olah2009multiple}.

The light curve amplitude $A$ has varied between 0.00 and 0.24 mag. The short-term variations are quite regular, but do not show any suggestive signs of cyclic behaviour. In particular, correlation between the independent $M$ and $A$ estimates is absent with the linear Pearson correlation coefficient being $r=0.10$.

When the amplitude is at its lowest, the data are best described by a constant
brightness model. This indicates that occasionally the light curve of \object{LQ~Hya} reduces to such low amplitudes that the periodic variability is buried under the random observational errors and systematic errors introduced by the model. Physically this means a nearly total absence of spots or alternatively an axisymmetric spot distribution.

For \object{LQ~Hya} we find two epochs with repeated constant brightness models, first during the years 1989--90 and later during 1996--98. During the first of these epochs the brightness of the star was sharply rising from a deep minimum (see Fig. \ref{cyc}). This means that there must have been a varying amount of starspots present and the constant brightness models are best understood as times of axisymmetric spot coverage. During the latter of these epochs the mean brightness of the star was near its maximum indicating a relative lack of spots. On the other hand, the maximum values of $A$ above 0.2 mag are a substantial indication of strongly concentrated spot activity.

The variability of the light curve period $P$ seems to consist of random fluctuations. These could be caused by starspots randomly occurring on different latitudes having different rotation periods and thus tracing differential rotation. Alternatively they may be caused by changes in the light curve shape as some active regions decay and others form at different longitudes. Following the assumption that the random fluctuations are caused by differential rotation, they are used to give an estimate for it in Sect. \ref{diffrot}.

\subsection{Differential rotation}
\label{diffrot}

An estimate for the differential rotation of \object{LQ~Hya} was obtained from period fluctuations using the formula \citep{jetsu1993decade}

\begin{equation}
Z = \frac{6\Delta P_{\rm w}}{P_{\rm w}},
\end{equation}

\noindent where $w_i=\sigma_{P,i}^{-2}$ are the weights, $P_{\rm w} = \sum{w_iP_i}/\sum{w_i}$ is the weighted mean and $\Delta P_{\rm w} = \sqrt{\sum{w_i(P_i-P_{\rm w})^2}/\sum{w_i}}$ is the weighted standard deviation. The parameter $Z$ measures the relative variability of the light curve period within its $\pm 3 \sigma$ limits.

When estimating differential rotation from photometric period fluctuations, the range of period fluctuations is often equated directly with the absolute value of the differential rotation coefficient $k$. This can generally not be assumed. The period fluctuations can provide information of the rotation period only from that latitude range of the stellar surface on which the spot activity occurs. As this range might be quite limited, we do not expect the photospheric period fluctuations to correspond to the total range of photospheric rotation periods present on the star. However, in the absence of any knowledge about the true latitude range of the spot activity, we may expect $|k|$ to be of the order of $Z$ or somewhat greater. Note that the sign of $k$ remains undetermined from photometry alone.

For \object{LQ~Hya} we get $P_{\rm w} \pm \Delta P_{\rm w} = 1.6043 \pm 0.0052$ d, corresponding to $|k| \approx Z = 0.020$ or $\Delta\Omega \approx 0.078$ rad/d. This is in line with previous estimates, especially those by \citet{jetsu1993decade} and \citet{you2007photometric}, who estimated the differential rotation in a similar manner.

We add a few caveats to the interpretation of the period fluctuations as tracers of differential rotation. First, if the starspots are caused by a large scale dynamo field, it is possible that they do not follow the surface rotation of the star \citep{korhonen2011}. It is also possible that the small period fluctuations are not caused by starspots having different rotation periods around the star but rather due to starspot growth and decay affecting the light curve shape. Lastly, we note that in datasets that have few data points and a low light curve amplitude there is considerable uncertainty in the period detection \citep[Table 2 in][]{lehtinen2011continuous}.

\subsection{Minimum phases $\phi_{\rm min}$ and active longitudes}
\label{alsect}

\begin{figure*}
\centering
\includegraphics[width=17cm]{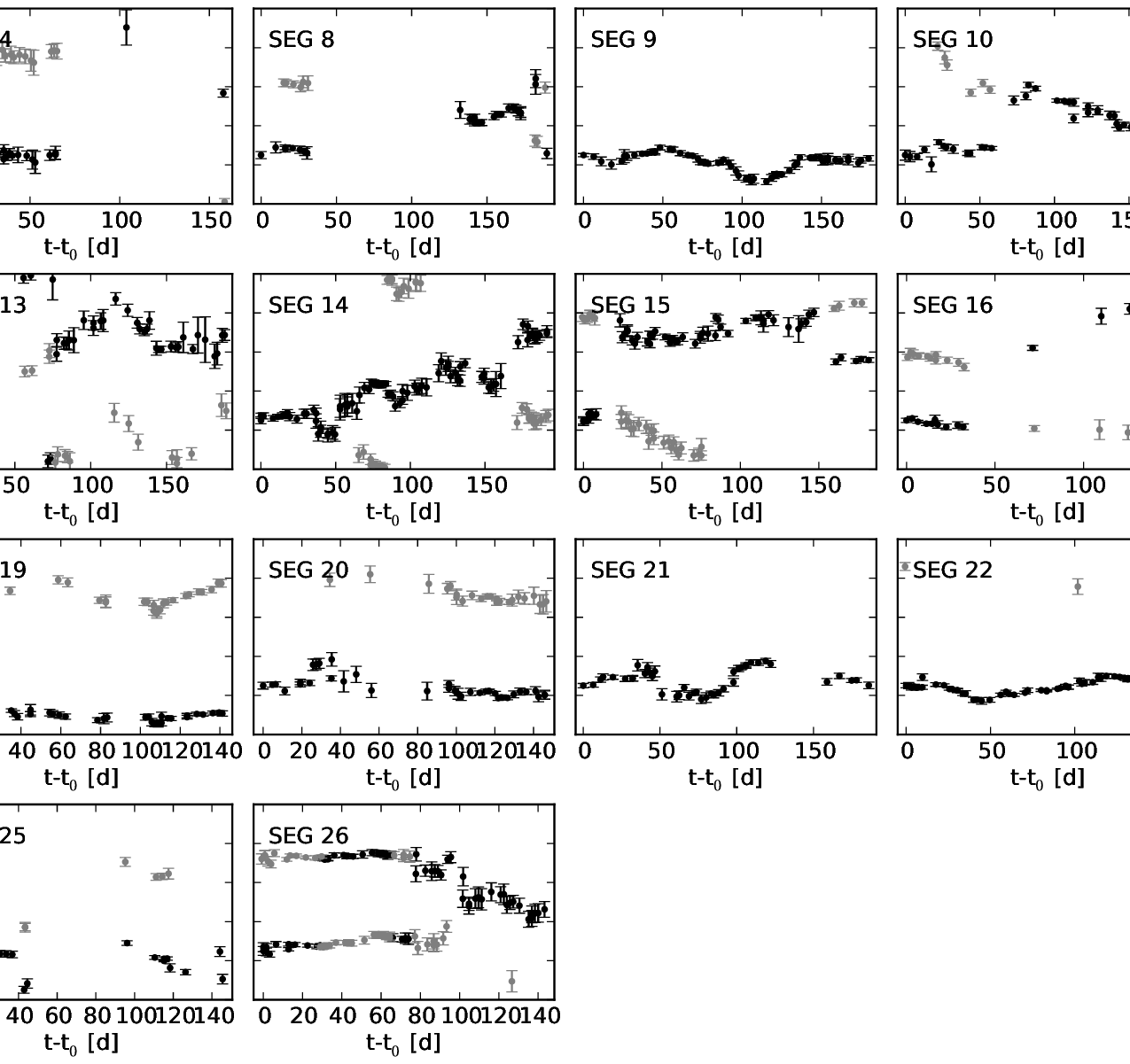}
\caption{Primary (black) and secondary (grey) light curve minimum phases, $\phi_{\rm min,1}$ and $\phi_{\rm min,2}$, for segments that have $n_{\rm rel} \geq 10$ datasets with reliable light curve models. The minimum phases are computed from the primary and secondary light curve minimum epochs, $t_{\rm min,1}$ and $t_{\rm min,2}$, with the median light curve periods $P_{\rm med}$ of each segment and the first primary minimum set at $\phi=0.25$.}
\label{alseg}
\end{figure*}

The longitudinal distribution of the spots on \object{LQ~Hya} can be studied using the epochs for the primary and secondary minima $t_{\rm min,1}$ and $t_{\rm min,2}$ of the modelled light curve. For any reasonable rotation period estimate $P$, these can be transformed into light curve minimum phases

\begin{equation}
\phi_{\rm min} = {\rm FRAC}\left[\frac{t_{\rm min}}{P}\right],
\end{equation}

\noindent where ${\rm FRAC}[x]$ removes the integer part of $x$.

The minimum phases $\phi_{\rm min,1}$ and $\phi_{\rm min,2}$ for individual segments with $n_{\rm rel} \geq 10$ datasets with reliable parameter estimates are presented in Fig. \ref{alseg}. The phases are folded from the primary and secondary minimum epochs using the median photometric period $P_{\rm med}$ of each segment and placing the first primary minimum of each segment at $\phi=0.25$. For each segment, time is given starting from the the first analysed data point in the segment as given in Table \ref{segtab}.

A recurring pattern between the segments has been the presence of one or two active regions wandering slightly in phase (Fig. \ref{alseg}). In many segments the two active regions inferred from the light curve minima stayed roughly at the opposite sides of the star resembling long-lived active longitudes. In other segments, such as SEG 14 and SEG 26, there are clear examples of one light curve minimum splitting into two or two minima merging into one as the underlying active regions have moved away or towards each other. At some critical phase separation the two active areas have moved too close to each other to produce separate light curve minima and instead we only observe one merged minimum \citep[Eq. 12]{lehtinen2011continuous}. In several segments (SEG 3,10, 15, 16 and 26) the main photometric minimum switched from one phase to another. Although occurring only on a short timescale, this may be analogous to the flip-flop behavior discovered in other stars \citep{jetsu1993spot}. This phenomenon is observed as a sudden change of the spot activity to the opposite side of the star. The original active area may survive the flip-flop with weakened level of activity or it may disappear completely. The examples that we find for \object{LQ~Hya} resemble the former of these types.

To examine whether there have been any long lasting active longitudes on \object{LQ~Hya}, i.e. that the light curve minima have clustered around certain phases, we performed the Kuiper test \citep{kuiper1960tests} for the minimum epochs $t_{\rm min}$. Our formulation for the test is from \citet{jetsu1996searching}. This procedure computes the Kuiper periodogram for $t_{\rm min}$ and tests the null hypothesis of uniform (i.e. random) phase distribution. It also determines the critical levels for the most significant periods. Examples of application of this method can be found in \citet{jetsu1996active} and \citet{lehtinen2011continuous}.

\begin{figure}
\resizebox{\hsize}{!}{\includegraphics{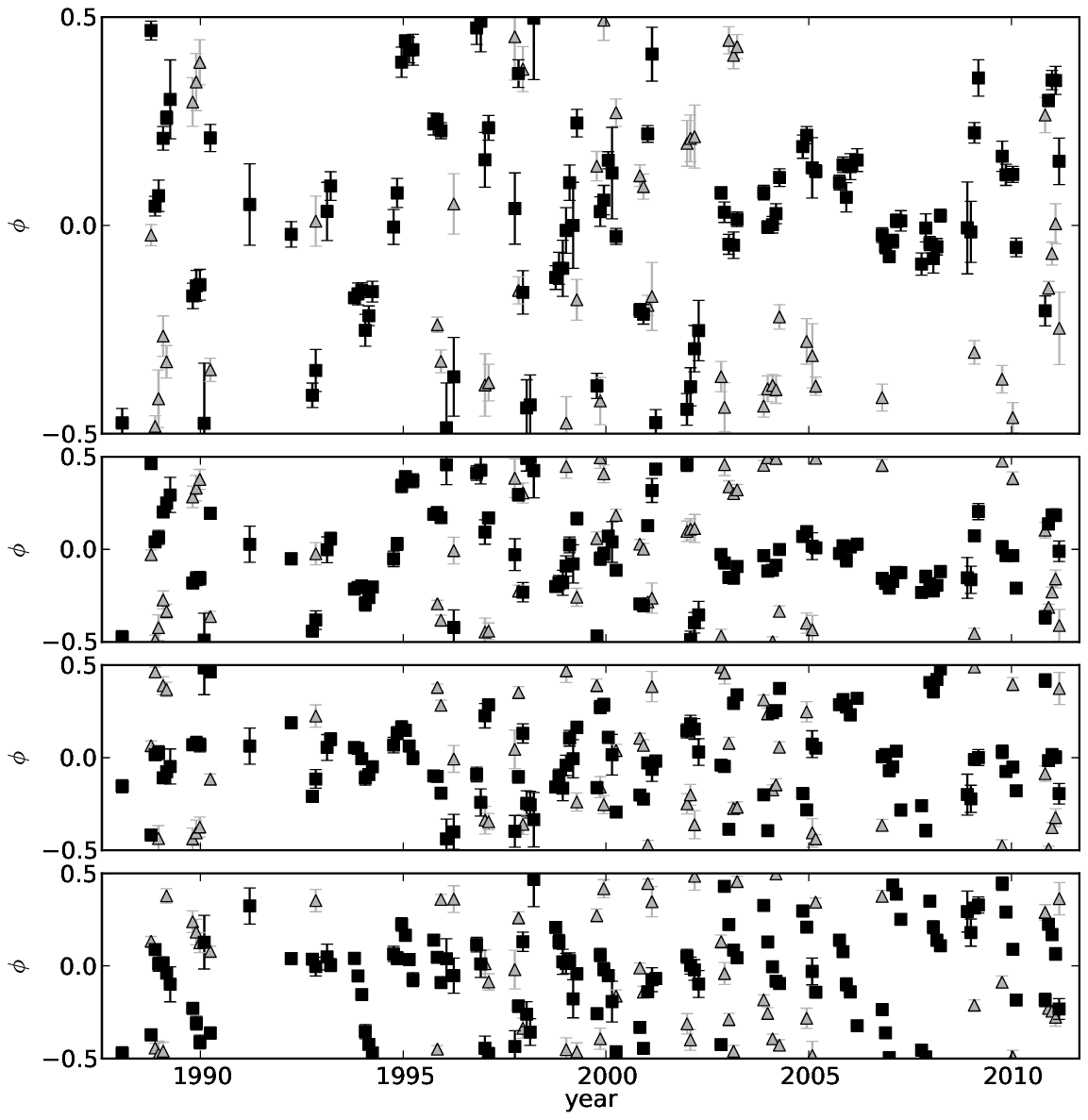}}
\caption{Light curve minimum phases folded according to the ephemeris ${\rm HJD_{min}} = 2447201.3 + 1.60369E$ for all of the minimum epochs (\textit{top}) and according to the seasonally determined ephemerides: ${\rm HJD_{min}} = 2447201.3 + 1.60374E$ \textit{second from top}, ${\rm HJD_{min}} = 2447200.8 + 1.68929E$ \textit{second from bottom} and ${\rm HJD_{min}} = 2447201.3 + 1.61208E$ \textit{bottom}. Black squares denote primary minima and grey triangles secondary minima.}
\label{al}
\end{figure}

The Kuiper test was performed for all reliable primary light curve minimum epochs from the independent (i.e. non-overlapping) datasets. This set comprised of 111 epochs. The most significant period found for all of the epochs was $P_{\rm al}=1.603693\pm0.000058$~d with the critical level $Q=6.1\cdot10^{-5}$. All independent minima, folded into minimum phases $\phi_{\rm min}$ with this period, are presented in the top panel of Fig. \ref{al}. The structure responsible for this period is clearly visible between the years 2003 and 2009. During this time interval the primary light curve minima were confined within a tight phase region. There are no signs of this structure being extant either before 2003 or after 2009, where the minimum phases show no structure with this period. An approximate ephemeris for the central meridian passing of this active longitude is given by

\begin{equation}
{\rm HJD_{min}} = 2447201.3 + 1.60369E.
\label{ephem}
\end{equation}

\noindent Even between the years 2003 and 2009 the active longitude has not remained stable but has moved back and forth within a phase range of $\Delta\phi\approx0.2$. A secondary active longitude, consisting solely of secondary light curve minima, appears to have been present for some years after 2003 with a phase separation of $\Delta\phi\approx0.5$ from the primary active longitude.

We performed additional Kuiper tests for the light curve minimum epochs to investigate the minimum phase distribution before and after the emergence of the obvious active longitude during 2003. The analysis was done for two samples of independent reliable minimum epoch estimates. The first sample consisted of the 67 independent primary minimum epochs in segments SEG~1 -- SEG~17 (i.e. until April 2002), while the second sample contained the remaining 44 independent primary minimum epochs in segments SEG~18 -- SEG~26 (i.e. starting from November 2002).

For the latter part of the minimum epochs, the analysis found the period $P_{\rm al,18-26}=1.60374\pm0.00013$ with the critical level $Q=2.3\cdot10^{-9}$. This corresponds to the same periodicity than $P_{\rm al}$ found for all minima but now with a significantly lower critical level. The minimum phases folded with this period (ephemeris ${\rm HJD_{min}} = 2447201.3 + 1.60374E$) are shown in the second from top panel of Fig. \ref{al} and follow very closely the pattern of the minimum phases folded with the period $P_{\rm al}$.

For the first part of the minimum epochs the analysis indicated $P_{\rm al,1-17}=1.68929\pm0.00008$ d to be the most significant periodicity with the critical level $Q=7.7\cdot10^{-5}$. The minimum phases folded with this period (ephemeris ${\rm HJD_{min}} = 2447200.8 + 1.68929E$) are presented in the second from bottom panel of Fig. \ref{al}. This folding brings out some structure for the primary minimum phases during the year 1995 and some time before that. However, the period $P_{\rm al,1-17}$ is considerably longer than either of the other two active longitude period estimates $P_{\rm al}$ and $P_{\rm al,18-26}$ or even the highest individual light curve period estimate $P_{\rm max}=1.6297$ d. We identify this to be a spuriois period.

The Kuiper periodogram for the segments SEG~1 -- SEG~17 also had a peak at $P_{\rm al,1-17}=1.61208\pm0.00008$ d closer to the other estimated periods related to the rotation of \object{LQ~Hya}. It is thus more likely to correspond to a physical phase structure. The minimum phases folded with this period (ephemeris ${\rm HJD_{min}} = 2447201.3 + 1.61208E$) are presented in the bottom panel of Fig. \ref{al}. However, the critical level $Q=1.1\cdot10^{-3}$ of this period is much higher than those of the other periods discussed in this section rendering the evidence for any coherent active longitudes before the end of 2002 uncertain.

The rest of the minimum epochs seem to exhibit no periodic structure at all.

\subsection{Time scale of change}

In addition to determining the light curve parameter estimates, the CPS also determines the time scale of change $T_{\rm C}$ for each individual dataset. This is defined as the time during which each model fit remains an adequate description for the subsequent datasets. As we have demonstrated \citep{lehtinen2011continuous}, the value of $T_{\rm C}$ can change dramatically from one dataset to another. A more meaningful value is the mean time scale of change $\overline{T}_{\rm C}$, which is just the mean of all $T_{\rm C}$.

For \object{LQ~Hya} we get $\overline{T}_{\rm C}=50.5$ d based on the individual $T_{\rm C}$ values of all independent datasets. This is longer than the maximum length of the datasets $\Delta T_{\rm max}=30.4$ d by a wide margin. In other words, the light curve of \object{LQ~Hya} typically retains its shape during the whole dataset and our choice for the dataset length is well justified.

As the value of $\overline{T}_{\rm C}$ estimates the typical time scale associated with the evolution of the spot configuration, it is interesting to check how this compares with the theoretical convective turnover time $\tau_c$. Previously, we used the interpolation formula by \citet{ossendrijver1997cycle}

\begin{equation}
\tau_c = -68.3 + 224.8(B-V) - 177.2(B-V)^2 + 57.0(B-V)^3,
\label{tauc}
\end{equation}

\noindent to estimate the convective turnover time for the young solar analogue \object{HD~116956} \citep{lehtinen2011continuous}. This formula is based on the theoretical calculations of \citet{kim1996theoretical}. By using the Hipparcos value $B-V=0.933$ for \object{LQ~Hya} \citep{perryman1997hipparcos}, we get $\tau_c=33.5$ d. Similarly as in the case of \object{HD~116956}, the values of $\overline{T}_{\rm C}$ and $\tau_c$ are of the same order of magnitude, $\tau_c$ being about 65\% of the length of $\overline{T}_{\rm C}$.

\section{Conclusions}

We have analysed 24 years of Johnson V-band photometry of the magnetically active star \object{LQ~Hya} with our recently published CPS method \citep{lehtinen2011continuous}. This method models the observed photometry with a variable order Fourier series using a sliding window for choosing the analysed datasets. The modelling provides estimates for the light curve mean differential magnitude $M$, total light curve amplitude $A$, photometric period $P$ and primary and secondary light curve minimum epochs $t_{\rm min,1}$ and $t_{\rm min,2}$ as functions of time.

\object{LQ~Hya} has displayed regular variability in the $M$ and $A$ estimates. Especially the variability of the mean differential magnitude $M$ resembles a segment from a quasi periodic time series. The variability could be explained with a roughly 13 year spot cycle. This is undoubtably the same structure that is behind the 11.4 yr cycle reported by \citet{olah2000multiperiodic}, the 15 yr cycle reported by \citet{berdyugina2002magnetic} and the 13.8 yr cycle suggested by \citet{kovari2004doppler}. However, the data analysed in this paper only includes one and a half cycles of this suggested activity cycle. Other studies \citep[e.g.][]{berdyugina2002magnetic,olah2009multiple} have included photometry going back to the end of 1982 which seems to qualitatively support the existence of the 13 yr cycle. But even this length of data is not enough to conclude whether the mean magnitude variations are indeed periodic in the long run.

We estimated the relative scale of photometric period variations within their $3\sigma$ limits to be $Z=0.020$ for \object{LQ~Hya}. Assuming that these variations are caused by starspots on different stellar latitudes experiencing differential rotation, we derive an estimate $|k|\approx0.02$ for the magnitude of the differential rotation coefficient or equivalently $\Delta\Omega \approx 0.078$ rad/d for the surface shear. This value indicates an almost rigid rotator and is virtually the same as the estimates by \citet{jetsu1993decade} and \citet{you2007photometric}. Alternatively it is possible that the period variations are caused by factors unrelated to surface differential rotation such as active region growth and decay or migration governed by the underlying dynamo field \citep{korhonen2011} and that the signal from any weak differential rotation gets buried under these. In either case our result conforms to theoretical results which indicate that fast rotating stars should approach rigid rotators \citep{kitchatinov1999differential}.

On the basis of the light curve minima, which indicate the main spot regions on the star, the typical configuration of the spot activity on \object{LQ~Hya} seems to be one or two active regions at different longitudes. These active regions seem to be relatively stable in time scales up to six months, although they usually undergo longitudinal migration during that time. Quite often the two active regions have moved so close to each other that we only observe one merged light curve minimum. Thus, the existence of even a third major active region producing a merged light curve minimum with either of the other two regions remains possible.

Contrary to the relatively short term stability of the individual active regions, there have been remarkably little stable structures in the long term phase distribution of the light curve minima. We found one active longitude, with a possible secondary active longitude at $\Delta\phi=0.5$ phase separation from the primary rotating with the period $P_{\rm al}=1.603693\pm0.000058$ d. However, this active longitude has only been present in the data between the years 2003 and 2009. Before this we found another possible structure around 1995 rotating with the period $P_{\rm al}=1.61208\pm0.00008$ d. This period detection was, however, not very significant. It is thus uncertain whether it corresponds to a real phase structure on the star or not. Contrary to \citet{berdyugina2002magnetic} we find no evidence for persistent active longitudes on \object{LQ~Hya} and no signs of regular flip-flop events.

We note that $P_{\rm al}=1.603693$ d is close to the mean photometric period $P_{\rm w}=1.6043$ d and well within the fluctuation of the individual light curve period estimates. In another active star \object{II~Peg} there were indications that the active longitude rotated faster than the star itself \citep{hackman2012iipeg,lindborg2011iipeg}, which could be explained by a dynamo wave propagating in the azimuthal direction \citep[e.g.][]{krause1980meanfield}. There does not seem to be such a dynamo wave on \object{LQ~Hya}.

The mean timescale of change for the light curve of \object{LQ~Hya} is $\overline{T}_{\rm C}=50.5$ d. This is longer than the maximum length of the individual datasets $\Delta T_{\rm max}$, which means that the light curve typically stays stable within the individual datasets. As a comparison, we estimated the convective turnover time to be $\tau_c=33.5$ d using the formula of \citet{ossendrijver1997cycle} (our Eq. \ref{tauc}). These two values are of the same order of magnitude. Previously we estimated a very similar ratio for these values, i.e. $\overline{T}_{\rm C}=44.1$ d and $\tau_c=28.5$ d, for the young solar analogue \object{HD~116956} \citep{lehtinen2011continuous}. Although it is not clear whether there is an actual connection between the two values, or if their ratio is simply governed by the choices made in the numerical procedure, such a result could be expected. After all, the strength of convection in the star is bound to be a strong factor in the evolution of the spot structures on late-type stars.

\begin{acknowledgements}
This work has made use of the SIMBAD data base at CDS, Strasbourg, France and NASA's Astrophysics Data System (ADS) bibliographic services. The work by J.L. was supported by the Finnish Graduate School in Astronomy and Space Physics. The work of T.H. was financed by the research programme ``Active Suns'' at the University of Helsinki. The work by P.K. was supported by the Vilho, Yrj\"{o} and Kalle V\"{a}is\"{a}l\"{a} Foundation. The automated astronomy program at Tennessee State University has been supported by NASA, NSF, TSU and the State of Tennessee through the Centers of Excellence program.
\end{acknowledgements}

\bibliographystyle{aa}
\bibliography{lqhya2012}

\begin{thebibliography}{35}
\expandafter\ifx\csname natexlab\endcsname\relax\def\natexlab#1{#1}\fi

\bibitem[{Berdyugina {et~al.}(2002)Berdyugina, Pelt, \&
  Tuominen}]{berdyugina2002magnetic}
Berdyugina, S., Pelt, J., \& Tuominen, I. 2002, \aap, 394, 505

\bibitem[{Donati(1999)}]{donati1999magnetic}
Donati, J.-F. 1999, \mnras, 302, 457

\bibitem[{Donati {et~al.}(2003)Donati, Collier~Cameron, Semel, Hussain, Petit,
  Carter, Marsden, Megnel, L\'{o}pez~Ariste, Jeffers, \&
  Rees}]{donati2003dynamo}
Donati, J.-F., Collier~Cameron, A., Semel, M., {et~al.} 2003, \mnras, 345, 1145

\bibitem[{Eggen(1984{\natexlab{a}})}]{eggen1984systematic}
Eggen, O. 1984{\natexlab{a}}, \aj, 89, 1358

\bibitem[{Eggen(1984{\natexlab{b}})}]{eggen1984a0}
Eggen, O. 1984{\natexlab{b}}, \apjs, 55, 597

\bibitem[{Eggen(1989)}]{eggen1989large}
Eggen, O. 1989, \pasp, 101, 366

\bibitem[{Fekel {et~al.}(1986{\natexlab{a}})Fekel, Bopp, Africano, Goodrich,
  Palmer, Quingley, \& Simon}]{fekel1986chromospherically}
Fekel, F., Bopp, B., Africano, J., {et~al.} 1986{\natexlab{a}}, \aj, 92, 1150

\bibitem[{Fekel \& Henry(2005)}]{fekel2005chromospherically}
Fekel, F. \& Henry, G. 2005, \aj, 129, 1669

\bibitem[{Fekel {et~al.}(1986{\natexlab{b}})Fekel, Moffett, \&
  Henry}]{fekel1986survey}
Fekel, F., Moffett, T., \& Henry, G. 1986{\natexlab{b}}, \apjs, 60, 551

\bibitem[{Hackman {et~al.}(2012)Hackman, Mantere, Lindborg, Ilyin, Kochukhov,
  Piskunov, \& Tuominen}]{hackman2012iipeg}
Hackman, T., Mantere, M., Lindborg, M., {et~al.} 2012, \aap, 538, A126

\bibitem[{Henry(1995)}]{henry1995development}
Henry, G. 1995, in Astronomical Society of the Pacific Conference Series,
  Vol.~79, Robotic Telescopes. Current Capabilities, Present Developments, and
  Future Prospects for Automated Astronomy, ed. G.~Henry \& J.~Eaton, 44--64

\bibitem[{Henry {et~al.}(1996)Henry, Soderblom, Donahue, \&
  Baliunas}]{henry1996survey}
Henry, T., Soderblom, D., Donahue, R., \& Baliunas, S. 1996, \aj, 111, 439

\bibitem[{Jetsu(1993)}]{jetsu1993decade}
Jetsu, L. 1993, \aap, 276, 345

\bibitem[{Jetsu(1996)}]{jetsu1996active}
Jetsu, L. 1996, \aap, 314, 153

\bibitem[{Jetsu \& Pelt(1996)}]{jetsu1996searching}
Jetsu, L. \& Pelt, J. 1996, \aaps, 118, 587

\bibitem[{Jetsu {et~al.}(1993)Jetsu, Pelt, \& Tuominen}]{jetsu1993spot}
Jetsu, L., Pelt, J., \& Tuominen, I. 1993, \aap, 278, 449

\bibitem[{K\H{o}v\'{a}ri {et~al.}(2004)K\H{o}v\'{a}ri, Strassmeier, Granzer,
  Weber, Ol\'{a}h, \& Rice}]{kovari2004doppler}
K\H{o}v\'{a}ri, Z., Strassmeier, K., Granzer, T., {et~al.} 2004, \aap, 417,
  1047

\bibitem[{Kim \& Demarque(1996)}]{kim1996theoretical}
Kim, Y.-C. \& Demarque, P. 1996, \apj, 457, 340

\bibitem[{Kitchatinov \& Olemskoy(2011)}]{kitchatinov2011differential}
Kitchatinov, L. \& Olemskoy, S. 2011, \mnras, 411, 1059

\bibitem[{Kitchatinov \& R\"{u}diger(1999)}]{kitchatinov1999differential}
Kitchatinov, L. \& R\"{u}diger, G. 1999, \aap, 344, 911

\bibitem[{Korhonen \& Elstner(2011)}]{korhonen2011}
Korhonen, H. \& Elstner, D. 2011, \aap, 532, A106

\bibitem[{Krause \& R\"{a}dler(1980)}]{krause1980meanfield}
Krause, F. \& R\"{a}dler, K.-H. 1980, {Mean-field magnetohydrodynamics and
  dynamo theory} (Oxford: Pergamon Press)

\bibitem[{Kuiper(1960)}]{kuiper1960tests}
Kuiper, N. 1960, in Proc. Koningkl. Nederl. Akad. Van Wettenschappen, Ser. A,
  Vol.~63, 38--47

\bibitem[{Lehtinen {et~al.}(2011)Lehtinen, Jetsu, Hackman, Kajatkari, \&
  Henry}]{lehtinen2011continuous}
Lehtinen, J., Jetsu, L., Hackman, T., Kajatkari, P., \& Henry, G. 2011, \aap,
  527, A136

\bibitem[{Lindborg {et~al.}(2011)Lindborg, Korpi, Hackman, Tuominen, Ilyin, \&
  Piskunov}]{lindborg2011iipeg}
Lindborg, M., Korpi, M., Hackman, T., {et~al.} 2011, \aap, 526, A44

\bibitem[{Montes {et~al.}(2001)Montes, L\'{o}pez-Santiago, G\'{a}lvez,
  Fern\'{a}ndez-Figueroa, De~Castro, \& Cornide}]{montes2001late}
Montes, D., L\'{o}pez-Santiago, J., G\'{a}lvez, M., {et~al.} 2001, \mnras, 328,
  45

\bibitem[{Nakajima \& Morino(2012)}]{nakajima2012potential}
Nakajima, T. \& Morino, J.-I. 2012, \aj, 143, 1

\bibitem[{Ol\'{a}h {et~al.}(2009)Ol\'{a}h, Koll\'{a}th, Granzer, Lanza,
  J\"{a}rvinen, Korhonen, Baliunas, Soon, Messina, \&
  Cutispoto}]{olah2009multiple}
Ol\'{a}h, K., Koll\'{a}th, Z., Granzer, T. amd~Strassmeier, K., {et~al.} 2009,
  \aap, 501, 703

\bibitem[{Ol\'{a}h {et~al.}(2000)Ol\'{a}h, Koll\'{a}th, \&
  Strassmeier}]{olah2000multiperiodic}
Ol\'{a}h, K., Koll\'{a}th, Z., \& Strassmeier, K.~G. 2000, \aap, 356, 643

\bibitem[{Ossendrijver(1997)}]{ossendrijver1997cycle}
Ossendrijver, A. 1997, \aap, 323, 151

\bibitem[{Perryman {et~al.}(1997)Perryman, Lindegren, Kovalevsky, H{\o}g,
  Bastian, Bernacca, Cr{\'e}z{\'e}, Donati, Grenon, van Leeuwen, van~der Marel,
  Mignard, Murray, Le~Poole, Schrijver, Turon, Arenou, Froeschl{\'e}, \&
  Petersen}]{perryman1997hipparcos}
Perryman, M., Lindegren, L., Kovalevsky, J., {et~al.} 1997, \aap, 323, L49

\bibitem[{Rice \& Strassmeier(1998)}]{rice1998doppler}
Rice, J. \& Strassmeier, K. 1998, \aap, 336, 972

\bibitem[{Strassmeier {et~al.}(1993)Strassmeier, Rice, Wehlau, Hill, \&
  Matthews}]{strassmeier1993surface}
Strassmeier, K., Rice, J., Wehlau, W., Hill, G., \& Matthews, J. 1993, \aap,
  268, 671

\bibitem[{White {et~al.}(2007)White, Gabor, \& Hillebrand}]{white2007high}
White, R., Gabor, J., \& Hillebrand, L. 2007, \aj, 133, 2524

\bibitem[{You(2007)}]{you2007photometric}
You, J. 2007, \aap, 475, 309

\end{thebibliography}

\end{document}